\documentclass[aps,preprintnumbers,twocolumn,amsmath,reprint,amssymb,nofootinbib,superscriptaddress,notitlepage]{revtex4-1}

\usepackage{slashed}
\usepackage{soul}
\usepackage{mathrsfs}
\usepackage{bm}
\usepackage{braket}
\usepackage{txfonts}
\usepackage{amssymb}
\usepackage{indentfirst}
\usepackage{hyperref}
\usepackage{graphicx,booktabs}
\usepackage{multirow}
\usepackage{overpic}
\usepackage{color}
\usepackage{amssymb}
\usepackage[norule]{footmisc}

\begin{document}
\title{Improved description of light nuclei through chiral effective field theory {{at leading order}}}

\author{M. {S\'anchez S\'anchez}}
\email{sanchez@cenbg.in2p3.fr}
\affiliation{CENBG (CNRS/IN2P3 -- Universit\'e de Bordeaux), 33175 Gradignan cedex, France}

\author{N. A. {Smirnova}}
\email{smirnova@cenbg.in2p3.fr}
\affiliation{CENBG (CNRS/IN2P3 -- Universit\'e de Bordeaux), 33175 Gradignan cedex, France}

\author{A. M. Shirokov}
\email{shirokov@nucl-th.sinp.msu.ru}
\affiliation{Department of Physics and Astronomy, Iowa State University, Ames, Iowa 50011}
\affiliation{Skobeltsyn Institute of Nuclear Physics, Lomonosov Moscow State University, Moscow 119991, Russia}
\affiliation{Pacific National University, 136 Tikhookeanskaya st., Khabarovsk 680035, Russia}

\author{P. Maris}
\email{pmaris@iastate.edu}
\affiliation{Department of Physics and Astronomy, Iowa State University, Ames, Iowa 50011}

\author{J. P. Vary}
\email{jvary@iastate.edu}
\affiliation{Department of Physics and Astronomy, Iowa State University, Ames, Iowa 50011}

\date{\today}
\begin{abstract}
\noindent We propose an arrangement of the most commonly invoked version of the two-nucleon chiral potential such that the low-lying amplitude zero of the $^1S_0$ partial wave is captured at leading order of the effective expansion. Adopting other partial waves from the LENPIC interaction, we show how this modification yields an improved description of ground-state energies and point-proton radii of three test nuclei. 
\end{abstract}

\pacs{13.60.Le, 12.39.Mk,13.25.Jx}

\maketitle

\section{Introduction}

One of the fundamental{ }challenges in nuclear physics{ }is to 
provide a consistent --- as well as phenomenologically successful --- derivation of the nuclear potential grounded on first principles. The nuclear effective field theory (EFT) program offers a way to address this challenge{. Here,} the link with the underlying quantum chromodynamics (QCD) leans on the fact that the effective Lagrangian fulfills all QCD symmetries --- most particularly the {spontaneously} and explicitly broken chiral symmetry, as in{ }chiral EFT ($\chi$EFT) \cite{VanKolckBedaque02,ModTheNucFor,Machleidt'n'Entem,StatPerp}{. The latter} aims at generalizing the scheme {of} chiral perturbation theory ($\chi$PT) \citep{Scherer03} to non-perturbative physics, namely nuclear systems. {The $\chi$EFT Lagrangian is written in terms of nucleons, pions and other hadron fields instead of the underlying quarks and gluons of QCD.} Since its symmetries are compatible with an infinite number of terms, it becomes mandatory to establish a hierarchical principle {(``power counting'')} that discriminates which terms should be used {for consistency} when computing {observables.  This enables one} to express the EFT predictions as{ }series in powers of {a} small parameter $Q/M_{\text{hi}}${. Here,} $Q$ ($M_{\text{hi}}$) {stands for} the magnitude of the typical external three-momentum of a process amenable to the EFT (the momentum scale at which the EFT breaks down and needs to be replaced by{ another }{theory} that underlies the former); in $\chi$EFT, this is of the order of the pion mass, $Q\sim 100\,\text{MeV}$ (the chiral-symmetry-breaking scale, $M_{\text{hi}}\sim 1\,\text{GeV}$). 

The initial applications of the nuclear EFT program {were} provided by pioneering studies in the early and middle 90s \citep{Weinberg90,Weinberg91,Rho:1990cf,Ordonez92,Ordonez94,Ordonez96}{. They} were grounded on the assumption that the nuclear potential and currents obey a power counting corresponding with that of $\chi$PT, called naive dimensional analysis (NDA) \cite{Manohar:1983md,Georgi93} or simply ``Weinberg power counting''. However, in non-perturbative physics this program {has been criticized for leading to} inconsistencies with the renormalization-group-invariance (RGI) (or cutoff-convergence) principle, {\textit{i.e.} for displaying model dependence,} see \textit{e.g.} Refs. \citep{KSW96,NTvK,PavonValderrama:2005uj}; {for a different interpretation in terms of a framework valid at a defined scale}, see \textit{e.g.} Refs. \citep{Epelbaum09,Epelbaum13,Epelbaum18,Epelbaum20}.

From a purely phenomenological point of view, it is worth noting that, at leading order (LO) in the expansion, the Weinberg counting fails to produce a qualitatively correct description of two-nucleon ($NN$) scattering {in the $^1S_0$ channel} at momenta $Q\sim m_\pi$ due to a lack of repulsion among the nucleons. This may be remedied through an enhancement of beyond-LO terms in the effective potential. As proposed in Ref. \citep{zero}, this enhancement is sufficient to reproduce the amplitude zero that shows up in this wave at a relatively soft scattering momentum. Consequently, the convergence of the effective expansion is improved.

In modern calculations, $\chi$EFT plays a leading role, chiral potentials being a basic ingredient for understanding nuclear structure and reactions with \textit{ab initio} methods (see Ref. \citep{abinitio} for an overview). Among such methods, one {of} the most{ }versatile is the no-core shell model (NCSM) \citep{Barrett13}. {In this approach,} the $A$-body non-relativistic Schr\"odinger equation is{ }solved {in a basis representation which is most often chosen to be} the spherical harmonic-oscillator (HO) basis{. All} the (structureless) nucleons are treated as active degrees of freedom, and the Slater-determinant expansion is built up from HO single-particle wavefunctions depending on the HO frequency{. This} allows {one} to reformulate the many-body problem as a symmetric {sparse} eigenvalue problem, whose solution has the finite size of the model space as its only source of uncertainty. {One} version of the chiral potential that is{ }employed in current NCSM calculations is essentially given by Weinberg power counting. However, the abovementioned lack of repulsion in the $^1S_0$ channel results in an overbinding pattern of light nuclei at LO (see {the work by the LENPIC Collaboration in Refs. }\citep{Maris:2016wrd,LENPIC18,Epelbaum:2018ogq}). The convergence of the expansion may be thus accelerated by means of a modification of the original prescription along the lines explored in Ref. \citep{zero}{. In particular, we replace} the bare, partial-wave projected  $NN$ potential of the $^1S_0$ wave (where the centrifugal suppression that appears for channels with $\ell\geqslant1$ \citep{PeripheralSinglets} is absent) and {we retain} the LO interactions present in all the remaining channels. This{ }produces a very significant improvement in the predictions for ground-state energies and point-proton RMS radii of three light nuclei, namely $^3$H, $^4$He and $^6$He, as {we show in the present work}.

This article is structured as follows. In Sec. \ref{SecI}, the {issues with} Weinberg power counting {in} the $^1S_0$ two-nucleon channel are examined, the strategy used here to improve on such problems is described, and the details of our calculation at the two-body level are provided. In Sec. \ref{SecII}, we show the LO results for the ground-state energies and radii of $^3$H, $^4$He and $^6$He both for the LO original LENPIC interaction (see Ref. \citep{Semilocal}) and for the LO modified one, examine the convergence of the energies in the {infinite-basis} limit, and argue that {the overestimation of binding energies and the underestimation of radii} coming from the LO original LENPIC potential are {significantly} alleviated after the modification we propose. Finally, in Sec. \ref{SecIII} we present our conclusions and {outline} our ideas for future research work.

\section{The two-nucleon $\mathbf{^1}$\textit{S}$\mathbf{_0}$ channel in Chiral EFT \label{SecI}}

The emergence of a pole in the $NN$ $^1S_0$ amplitude at imaginary momentum $k \approx i/a$ {(note that the $\hbar=c=1$ units are used all through this work)}, {where} $a = -23.7\,\text{fm} = -\,(8.3\,\text{MeV})^{-1}$ {is} the scattering length in the neutron-proton channel \begin{footnote}{Unlike the Coulomb repulsion in the two-proton channel, the effects of charge-independence and charge-symmetry breakings that are inherent to the strong force {have} been neglected in our first-order approach. We recall \citep{Miller90} that those two {phenomena} may be respectively {quantified} in terms of the $^1S_0$ neutron-neutron ($nn$), 
proton-proton ($pp$),
and neutron-proton ($np$)  
scattering lengths \citep{PWA,nnonline} as
\begin{eqnarray*}
\frac{a_{np}-(a_{nn}+a_{pp}^{\text{(strong)}})/{2}}{a_{np}} \approx 0.24\,; \quad
\frac{a_{nn}-a_{pp}^{\text{(strong)}}}{a_{np}} \approx 0.05\,.
\end{eqnarray*}
}\end{footnote}, has long been identified as a very shallow virtual state{. This,} together with the loosely bound deuteron, anticipates the non-perturbative nature of nuclear physics already in its simplest manifestation --- the two-nucleon system. {As a consequence, }the LO part of the $NN$ interaction {has to be} fully iterated when inserted in a dynamical equation (\textit{e.g.} Schr\"odinger) that governs the system. In particular, the attraction provided by the {one-pion-exchange (OPE)} contribution --- the main long-range effect --- to the $^1S_0$ potential {is relatively mild. This introduces an attractive short-range contribution also at LO, sufficient} to render the aforementioned almost-bound state. The momentum-space representation of this short-range part reads as a pure constant $C_0${. Its} inclusion as a first-order effect is grounded on two complementary arguments. On the one hand, this contact force comes from a four-nucleon vertex \textit{without} derivative nor pion-mass insertions; hence,  according to {NDA}, it will be parametrically enhanced by $\mathcal{O}(M_{\text{hi}}^n/Q^n)$ with respect to a diagram with $n$ of such insertions \footnote{Actually, for the particular case of the $^1S_0$ wave the four-nucleon diagram with no derivative and two pion-mass insertions, which gives rise to a pointlike $D_2m_\pi^2$ interaction, happens to break this rule and is nominally as relevant as the $C_0$ vertex \citep{KSW96}. This observation, however, remains inconsequential in the pure-nucleon sector, provided that the pion mass is treated as a constant.

}. On the other hand, from an RGI perspective, the cutoff-independent contact piece of the {$^1S_0$-projected} OPE as it emerges from the effective Lagrangian would pose an ill-defined solution unless such a piece is reabsorbed into the running coupling $C_0$.

There is, however, another relevant feature of the $^1S_0$ partial wave that was recognized early on{. This is} the fact that the $NN$ scattering amplitude changes from positive to negative
at{ }momentum $k\equiv k_0\approx340\,\text{MeV}${. 
It is worth {recalling} that this fact motivated the inclusion of a short-range repulsive core in some of the earliest phenomenological models of the $NN$ interaction (see \textit{e.g.} Refs. \citep{Greenberger61,Otsuki64}). From a more modern perspective, provided that the hard scale in $\chi$EFT, usually identified as the typical mass of the lightest non-Goldstone hadrons, {respects} $M_{\text{hi}} \sim 1\,\text{GeV}$, then one should identify{ }\footnote{ Such an assumption is in good agreement with the LO nature of OPE in the $^1S_0$ wave since, in terms of power counting, this relies on $M_{NN}\sim Q$, where $M_{NN}$ sets the inverse strength of OPE in this wave; recall that $k_0$ happens to be only $\sim15\%$ numerically larger than $M_{NN}$ (see Ref. \citep{Jaber?} for a different approach to this).} $k_0\sim Q${. Hence}, $\chi$EFT should be well convergent in the $k\sim k_0$ momentum region{ --- in other words, beyond-LO corrections should not {offset a significant deficiency in} the LO result. {In addition, it seems appealing to have a} LO interaction that provides a satisfactory description of the phenomenological scattering matrix on a qualitative level, \textit{i.e.} in its gross features --- for instance not only its poles, but also its eventual zeros and changes of sign. This {concept appears particularly reasonable} if one adheres {to} the idea that only the LO part of the potential should be treated non-perturbatively, while $\text{N}^\nu\text{LO}$ terms should start contributing at $\nu\text{th}$ order in distorted-wave perturbation theory, as argued in Refs. \citep{NTvK,Birse:2005um,Valderrama:2009ei,Valderrama:2011mv,Long:2011qx,Long:2011xw,Long:2012ve,Song:2016}.

{To accomplish the vision just described, one confronts the fact that} the attraction provided by the short-range term $C_0$ in the Weinberg scheme is too strong to capture the amplitude zero at any reasonable momentum. Actually, the LO Weinberg prediction for this channel is a phase shift that becomes approximately constant ($\sim60{\,^\circ}$) in the middle-range region ($k\gtrsim100\,\text{MeV}$ all the way up to the pion-production threshold) provided that a reasonably hard momentum cutoff is employed ($\Lambda\gtrsim500\,\text{MeV}$, where a more precise estimate depends on the chosen regularization prescription). In Ref. \citep{zero} a new formulation of the short-range part of the LO potential was proposed in order to subsume the amplitude zero. 

To see how this fact can be exploited here, start by considering the part of the $\chi$EFT Lagrangian relevant for the two-nucleon $^1S_0$ channel in the standard arrangement,
\begin{widetext}
\begin{equation} 
\mathcal{L}_\chi^{\,(\text{W})} \,=\,
\frac{1}{2} \left(\partial_\mu\bm{\pi\cdot}\partial^\mu\bm{\pi}
- m_\pi^2{\bm{\pi}}^2\right)
\,+\, N^\dagger\,\Big[i\partial_0 + \frac{\overset{\rightarrow}{\nabla}{\color{white}\!i}^2}{2m_N}
-\frac{g_A}{2f_\pi}\bm{\tau}\bm{\cdot}
(\overset{\rightarrow}{\sigma}\cdot\overset{\rightarrow}{\nabla})\,\bm{\pi}
\Big]\,N \,-\, C_0\left(N^T\bm{P}_{^1S_0}N\right)^\dagger\bm{\cdot}\left(N^T\bm{P}_{^1S_0}N\right)\,+\,\dots\,,\label{Lchi}
\end{equation}
\end{widetext}
where $\bm{\pi}$ and $N$ denote the pion isotriplet and nucleon isodoublet fields {with isospin-averaged masses $m_\pi=138.04\,\text{MeV}$ and $m_N=938.92\,\text{MeV}$}, $g_A = 1.26$ and $f_\pi = 92.4\,\text{MeV}$ are the axial-coupling and pion-decay constants, $\bm{P}_{^1S_0} = \sigma_2\bm{\tau}\tau_2/\!\sqrt{8}$ is the two-nucleon projector in terms of the Pauli matrices $\overset{\rightarrow}{\sigma}$ ($\bm{\tau}$) acting on spin (isospin) space, and the ellipsis stands for more complicated terms suppressed by negative powers of the breakdown scale. 
Applying the usual Feynman rules in momentum space, the $^1S_0$ partial-wave projected two-nucleon potential is obtained to be
\begin{equation}
V_\chi^{\,(\text{W})}(p',p) = C_0 + V_\pi(p',p)\,, \label{VchiW}
\end{equation}
where $p$ ($p'$) is the magnitude of the relative momentum of the incoming (outgoing) nucleons, while the long-range component of the interaction is
\begin{equation}
V_\pi(p',p) = \frac{1}{m_N}\int_{\,0}^{\,\infty}\text{d}r\,r^2j_0(p'r)\,U_\pi(r)\,j_0(pr)\,,
\end{equation}
$j_0(x)=x^{-1}\sin x\,$ being the zeroth-order spherical Bessel function of the first kind, and
\begin{equation}
U_\pi(r) = -\frac{m_\pi^3}{M_{NN}}\,Y(m_\pi r)\,,\quad M_{NN} = \frac{16\pi f_\pi^2}{g_A^2m_N}\,,\quad Y(x) = \frac{e^{-x}}{x}\,.
\end{equation}
Besides, $C_0$ has been redefined with respect to Eq. \eqref{Lchi} through $C_0\,+\,4\pi/(m_NM_{NN})\,\to\,C_0$. The off-shell scattering matrix is then non-perturbatively found by solving the S-wave projected Lippmann-Schwinger equation
\begin{equation}
T(p',p;k) = V(p',p)+\frac{2}{\pi}\int_{\,0}^{\,\infty}\!\mathrm{d}q\,\frac{V(p',q)\,q^2\,T(q,p;k)}{(k^2-q^2)/m_N+i0^+} \label{LS}
\end{equation}
{for $V(p',p)\equiv V_\chi^{(\text{W})}(p',p)$. However, this happens to be  singular, as one can see by the fact that the integral in Eq. \eqref{LS} is linearly divergent; thus, a regularization prescription must be used. To be consistent with our adoption of potentials in the remaining partial waves from Ref. \citep{Semilocal}, we will apply a non-local regulator for the short-range component of the interaction and a local regulator for the long-range one,
\begin{eqnarray}
C_0 \,\to\, f_{\text{S}}(\tfrac{p'}{\Lambda})\,C_0\,f_{\text{S}}(\tfrac{p}{\Lambda})\,,&\quad& f_{\text{S}}(x) = e^{-x^2}\,; \label{short}\\
U_\pi(r) \,\to\, U_\pi(r)\,f_{\text{L}}(\tfrac{r}{R})\,,&\quad& f_{\text{L}}(x)=(1-e^{-x^2})^{\,6}\,, \label{long}
\end{eqnarray}
where the coordinate and momentum cutoffs $R$ and $\Lambda$ verify $R\Lambda=2$, so that 
\begin{eqnarray}
\int_{\,-\infty}^{\,+\infty}\frac{\mathrm{d}^3k}{(2\pi)^3}\,e^{i\vec{k}\cdot\vec{r}}\,f_{\text{S}}(\tfrac{k}{\Lambda}) \,\,&\propto& \,\,f_{\text{S}}(\tfrac{r}{R})   \\
 && \notag
\end{eqnarray} 
is fulfilled. The non-perturbative phase shift is obtained from the on-shell scattering matrix,
\begin{equation}
\delta(k) = \frac{1}{2i}\log\,[1-2ikm_NT(k,k;k)]\,.
\end{equation}
The value of $C_0$ is found by imposing the renormalization condition 
\begin{equation}
\lim_{k\to0}k\cot\delta(k)=-\frac{1}{a}\,. \label{rencondold}
\end{equation} 
If one chooses $R=0.9\,\text{fm}$ ($\Lambda=439\,\text{MeV}$), then $C_0 = -\left(440\,\text{MeV}\right)^{-2}$.

The proposal of Ref. \citep{zero} is to remedy the excess of attraction of the interaction \eqref{VchiW} through resumming into LO subleading terms that are repulsive enough to render the amplitude zero. This is done by means of a reparametrization of Eq.~\eqref{Lchi} grounded on the introduction of two auxiliary ``dibaryon'' \citep{phi} fields $\bm{\phi}_1$ and $\bm{\phi}_2$ such that the effective Lagrangian becomes
\begin{widetext}
\begin{eqnarray} 
\mathcal{L}_\chi^{\,(2\bm{\phi})} &=&
\frac{1}{2} \left(\partial_\mu\bm{\pi\cdot}\partial^\mu\bm{\pi}
- m_\pi^2{\bm{\pi}}^2\right)
\,+\, N^\dagger\,\Big[i\partial_0 + \frac{\overset{\rightarrow}{\nabla}{\color{white}\!i}^2}{2m_N}
-\frac{g_A}{2f_\pi}\bm{\tau}\bm{\cdot}
(\overset{\rightarrow}{\sigma}\cdot\overset{\rightarrow}{\nabla})\,\bm{\pi}
\Big]\,N \notag\\ &&+ \sum_{j=1,2} \left\lbrace\, \bm{\phi}_j^\dagger\,\bm{\cdot}\,\Bigg[\mathsf{\Delta}_j+c_j\,\Big(i\partial_0+\frac{\overset{\rightarrow}{\nabla}{\color{white}\!i}^2}{4m_N}
\Big)\,\Bigg]\,\bm{\phi}_j\,-\sqrt{\frac{4\pi}{m_N}}\left(\bm{\phi}_j^\dagger\bm{\cdot}N^T\bm{P}_{^1S_0}N\,+\,\text{H.c.}\right)\,\right\rbrace+\dots\,,\label{Lchiphi}
\end{eqnarray}
\end{widetext}
where the two-dibaryon low-energy couplings (LECs) --- the residual masses $\mathsf{\Delta}_j$ and the kinetic factors $c_j$ --- admit expansions in powers of $Q/M_{\text{hi}}$; for notation simplicity, here we will abbreviate $\mathsf{\Delta}_j\equiv\mathsf{\Delta}_j^{[0]}$ and $c_j\equiv c_j^{[0]}$, the superscript $[0]$ referring to the LO contribution. The prescription $c_1\equiv0$ is adopted, giving rise to the potential \begin{equation}V_\chi^{\,(2\bm{\phi})}(p',p,k) \,=\, \frac{4\pi}{m_N}\left(\frac{1}{\mathsf{\Delta}_1}+\frac{1}{\mathsf{\Delta}_2+c_2k^2/m_N}\right) \,+\, V_\pi(p',p) \label{Vchi2phi}\end{equation}
($1/\mathsf{\Delta}_1\,+\,1/M_{NN}\,\to\,1/\mathsf{\Delta}_1$). Then one can fit to \citep{zero}
\begin{equation}
\lim_{k\to0}k\cot\delta(k)=-\frac{1}{a}\,,\quad \lim_{k\to0}\frac{\partial}{\partial k^2}k\cot\delta(k) = \frac{r_0}{2}\,,\quad \delta(k_0)=0\,, \label{rencond}\end{equation}
$r_0=2.7\,\text{fm}$ being the $^1S_0$ $np$ effective range. 
In Ref. \citep{zero} it is shown that Eq. \eqref{Vchi2phi} yields a surprisingly good description of the phenomenological $^1S_0$ phase shift \citep{nnonline} in the whole elastic regime. However, note that this interaction is energy dependent, which is often a drawback in calculations beyond the two-body sector --- in general, it is unclear how to define the pair energy on which the pair potential would depend.

In this work we adopt a heuristic approach by exploring a momentum-dependent interaction such that its on-shell version coincides with Eq. \eqref{Vchi2phi}. This can be done in terms of the introduction of an auxiliary isovector field
\begin{eqnarray}
\bm{\Phi} &=& N^T\,\Big[\,\gamma^2+\Big(\,\overset{\rightarrow}{\nabla}-\overset{\leftarrow}{\nabla}\,\Big)^2\,\Big]^{-\frac{1}{2}}\,\bm{P}_{^1S_0}N \notag\\
&=& N^T\,\Bigg[\,\frac{1}{\gamma} - \frac{1}{2\gamma^3}\,\Big(\,\overset{\rightarrow}{\nabla}-\overset{\leftarrow}{\nabla}\,\Big)^2+\dots\,\Bigg]\,\bm{P}_{^1S_0}N\,,
\end{eqnarray}
so that the effective Lagrangian becomes \citep{BW}
\begin{widetext}
\begin{eqnarray} 
\mathcal{L}_\chi^{(\bm{\Phi})} &=&
\frac{1}{2} \left(\partial_\mu\bm{\pi\cdot}\partial^\mu\bm{\pi}
- m_\pi^2{\bm{\pi}}^2\right)
+ N^\dagger\Big[i\partial_0 + \frac{\overset{\rightarrow}{\nabla}{\color{white}\!i}^2}{2m_N}
-\frac{g_A}{2f_\pi}\bm{\tau}\bm{\cdot}
(\overset{\rightarrow}{\sigma}\cdot\overset{\rightarrow}{\nabla})\bm{\pi}
\Big]N - \frac{4\pi}{m_N}\Big[\frac{\gamma^2}{\Delta_1}\bm{\Phi}^\dagger\bm{\cdot}\bm{\Phi}+\frac{1}{\Delta_2}\left(N^T\bm{P}_{^1S_0}N\right)^\dagger\bm{\cdot}\left(N^T\bm{P}_{^1S_0}N\right)\Big]+\dots\,,
\notag\\\label{LchiPhi}
\end{eqnarray}
\end{widetext}
giving rise to a separable-plus-constant short-range potential
\begin{eqnarray}
V_\chi^{\,(\bm{\Phi})}(p',p) &=& \frac{4\pi}{m_N}\left[\frac{1}{\Delta_1}\,\mathcal{F}(\tfrac{p'}{\gamma})\,\mathcal{F}(\tfrac{p}{\gamma})+\frac{1}{\Delta_2}\right] +V_\pi(p',p)\,,\notag \\\mathcal{F}(x)&=&(1+x^2)^{-\frac{1}{2}} \label{VchiPhi}
\end{eqnarray}
($1/\Delta_2\,+\,1/M_{NN}\,\to\,1/\Delta_2$), which is again supplemented by {Eq.} \eqref{rencond} \footnote{In the $M_{NN}\to\infty$ limit of the interaction \eqref{VchiPhi}, Eq. \eqref{LS} admits an analytic solution and one can see explicitly that the resulting on-shell amplitude coincides with the one arising from the $M_{NN}\to\infty$ version of the interaction \eqref{Vchi2phi}. Also, it is interesting to note that this ``pionless'' limit of Eq. \eqref{VchiPhi} is compatible with a positive effective range, thus {circumventing} the Wigner-bound issues \citep{Wigner55,Phillips97} of momentum-dependent contact potentials such as the ones explored in Ref. \citep{Phillips98}. This observation is consistent with the conclusions of Ref. \citep{Beck19}, provided that the structure in Eq. \eqref{VchiPhi} is seen as the infinite resummation of interaction terms that appear in standard pionless EFT, where the coefficients in front of those terms are fixed beforehand.

\medskip

}. {We solve Eq. }\eqref{LS}{ with $V(p',p)\equiv V_\chi^{(\Phi)}(p',p)$, by means of} a regularization strategy analogous to the one of Eqs. \eqref{short}--\eqref{long}, {for $R=0.9\,\text{fm}$.} Through the best fit to Eq. \eqref{rencond}, we find $\Delta_1=-58\,\text{MeV}$, $\Delta_2=96\,\text{MeV}$, $\gamma=476\,\text{MeV}$. 

{In Fig. }\ref{fig:delta}{ we {plot} the $^1S_0$ phase shifts arising from the potentials $V_\chi^{(\text{W})}(p',p)$ and $V_\chi^{(\bm{\Phi})}(p',p)$ incorporating the regularization prescription and the renormalization conditions detailed above, together with the partial-wave analysis of Ref. }\citep{PWA}{. Note that the best fit corresponding to the $V_\chi^{(\bm{\Phi})}(p',p)$ interaction yields a good reproduction of the phenomenological curve for momenta up to $\sim 300$ MeV, but fails to reproduce the amplitude zero at the correct location, shifting it $\sim10\%$ to the right. However, {this is a regulator artifact ---} increasing {slightly} the momentum cutoff $\Lambda$ would remedy this flaw {\citep{zero}}. {Besides, for larger cutoffs ($\Lambda\gtrsim M_{\text{hi}}$) the difference between the curves emerging from both potentials becomes greater (see Fig. 7 of Ref. }\citep{zero}{, where a sharp-cutoff regularization prescription is adopted).}

\begin{figure*}[t!]
\begin{center}
\includegraphics[scale=.5]{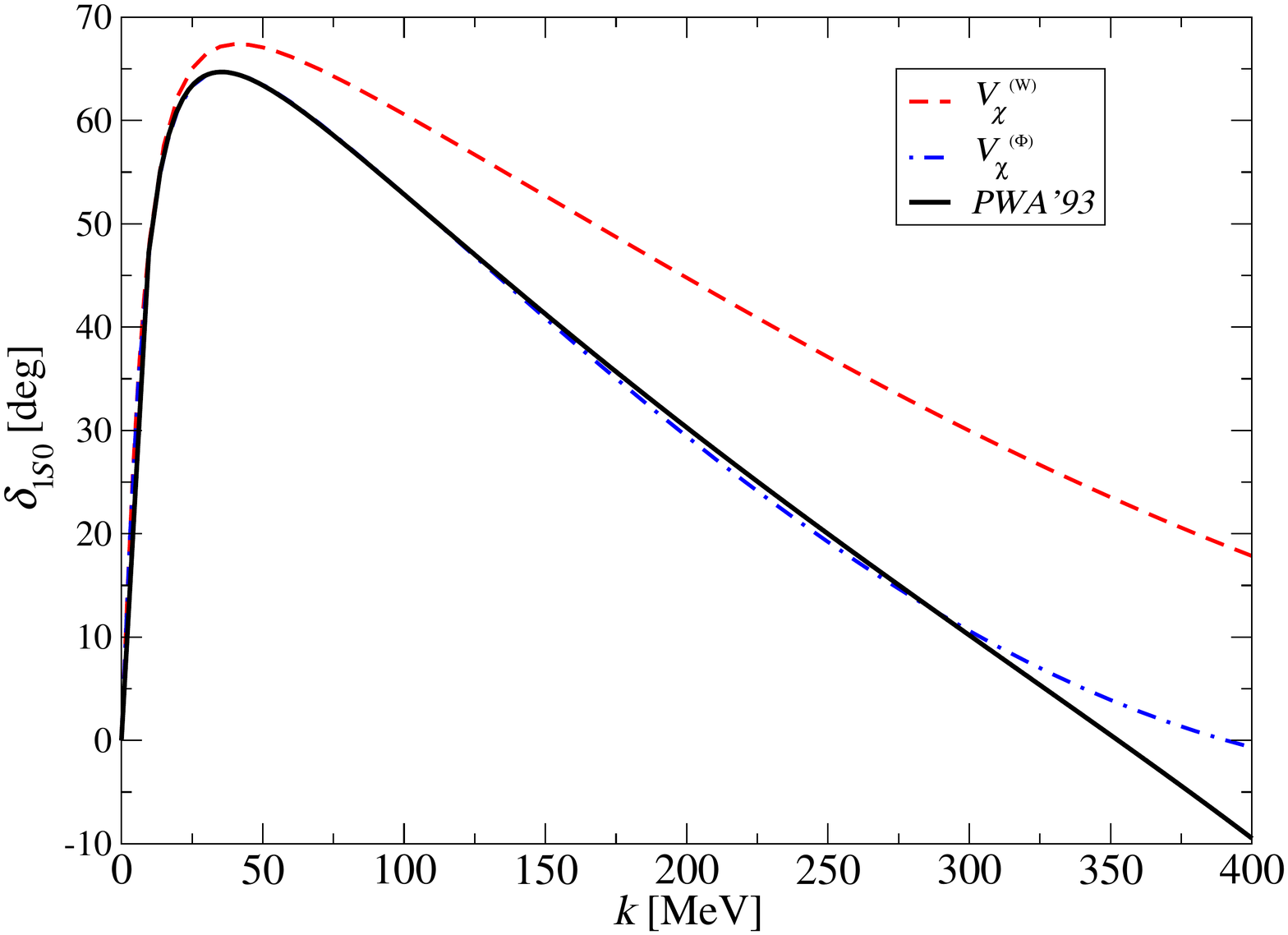}
\end{center}

\caption{$^1S_0$ phase shift (in degrees) as a function of the center-of-mass momentum (in MeV) for the potentials \eqref{VchiW} and \eqref{VchiPhi}, with the regularization prescriptions \eqref{short} and \eqref{long} ($R=0.9\,\text{fm}$) and the renormalization conditions \eqref{rencondold} and \eqref{rencond} respectively, depicted as dashed (red) and dotdashed (blue) curves. The solid (black) line is the phase shift extracted from the partial-wave analysis of Ref. \citep{PWA}.}
\label{fig:delta}
\end{figure*}

Once the LECs of Eq. \eqref{VchiPhi} are determined, the two-body matrix elements (TBMEs) of this interaction in the HO basis are found through sandwiching
between corresponding HO states $\ket{n,\ell}$:
\begin{widetext}
\begin{eqnarray}
\mathcal{V}_{n'n}^{\,(^1S_0)} &\equiv& \braket{n',0\,|\,V_\chi^{(\bm{\Phi})}\,|\,n,0} = \frac{2}{\pi}\int_0^\infty\mathrm{d}p'\,p'^2\,\mathrm{d}p\,p^2\,\psi_{n'0}^*(p')\,V_\chi^{\,(\bm{\Phi})}(p',p)\,\psi_{n0}(p) \notag\\
&=& \frac{1}{m_N}\left[8\int_0^\infty\mathrm{d}p'\,p'^2\,\mathrm{d}p\,p^2\,\psi_{n'0}^*(p')\,f_{\text{S}}(\tfrac{p'}{\Lambda})\left(\frac{1}{\Delta_1}\,\mathcal{F}(\tfrac{p'}{\gamma})\,\mathcal{F}(\tfrac{p}{\gamma})+\frac{1}{\Delta_2}\right)f_{\text{S}}(\tfrac{p}{\Lambda})\,\psi_{n0}(p)+\int_0^\infty\mathrm{d}r\,r^2\,\Psi_{n'0}^*(r)\,U_\pi(r)\,f_{\text{L}}(\tfrac{r}{R})\,\Psi_{n0}(r)\right]\,,\notag\\.
\end{eqnarray}
\end{widetext}
with $\psi_{n\ell}(p)$ and $\Psi_{n\ell}(r)$ the momentum- and coordinate-space representations of the radial basis functions at {radial quantum number} $n$ and orbital angular momentum $\ell$, given by \citep{Bansal18}
\begin{eqnarray}
\psi_{n\ell}(p) &=& (-1)^n\!\sqrt{\frac{2\Gamma(n+1)b^3}{\Gamma(n+\ell+\frac{3}{2})}}(pb)^\ell e^{-\frac{p^2b^2}{2}}L_n^{(\ell+\frac{1}{2})}(p^2b^2)\,; \notag \\
\Psi_{n\ell}(r) &=& \sqrt{\frac{2\Gamma(n+1)}{\Gamma(n+\ell+\frac{3}{2})b^3}}(r/b)^\ell e^{-\frac{r^2}{2b^2}}L_n^{(\ell+\frac{1}{2})}(r^2/b^2)\,, \label{eigenfunctions}
\end{eqnarray}
where $b=\sqrt{2/(m_N\Omega)}$ is the HO length, $\Omega$ being the HO frequency, and $L_n^{(\ell+\frac{1}{2})}$ is a generalized Laguerre polynomial. These TBMEs are then transformed to the single-particle basis, supplementing contributions in other partial waves from the LENPIC$^{[0]}$ two-nucleon interaction \citep{Maris:2016wrd,LENPIC18,Epelbaum:2018ogq} for implementing in our many-body calculations.

\section{Study of light nuclei \label{SecII}}

For our initial application to light nuclei we selectively investigate the ground-state properties of $^3$H, $^4$He and $^6$He using the NCSM approach. Our NCSM calculations have been carried out on the supercomputer \textit{Cori}, a Cray XC40 system at LBNL, by means of the highly parallelized nuclear-structure eigensolver known as \textit{Many-Fermion Dynamics for nuclei} (MFDn){ }\citep{Maris10,Aktulga14,Shao18}. 
The nuclear observables have been calculated as functions of the HO energy {$\hbar\Omega = (15, 17.5, \dots, \hbar\Omega_{\text{max}})\,\text{MeV}$}, where $\hbar\Omega_{\text{max}}$ has been chosen for each nucleus to provide a visual impression of the convergence (ranging from $\hbar\Omega_{\text{max}}=50\,\text{MeV}$ for the loosely bound $^6$He to $\hbar\Omega_{\text{max}}=70\,\text{MeV}$ for the tightly bound $^4$He). Our results have been obtained for different values of the parameter $N_{\text{max}} = 6, 8, \dots, 16$ (for $^3$H and $^4$He) or $N_{\text{max}} = 4, 6, \dots, 14$ (for $^6$He). These values of $N_{\text{max}}$ are both convenient and sufficient for our purposes. We recall that this parameter represents the maximum number of HO {excitation} quanta that can be shared among the $A$ nucleons above the minimum-energy configuration. Both $\hbar\Omega$ and $N_\text{max}${ }give a measure of the infrared and ultraviolet cutoffs, respectively, and fully determine the model space \citep{Coon12}{. {Convergence} is reached for $N_{\text{max}}\to\infty$, for which the results should become $\hbar\Omega$-independent. In the following, we will explain how we extrapolated our finite-$N_{\text{max}}$ results for both observables.

\subsection{{Ground-state} energies}

The results on ground-state energies of  $^3$H, $^4$He and $^6$He are shown in Fig. \ref{fig:1}: left panels present the results obtained with the original LENPIC$^{[0]}$ interaction \citep{Semilocal,Maris:2016wrd,LENPIC18,Epelbaum:2018ogq}, while right panels show the results obtained with the same interaction modified in the $^1S_0$ channel 
as described above.
We observe that the proposed modification removes much of the overbinding inherent to the conventional LO potential{. At the same time,} the convergence of the calculations
as a function of $N_{\text{max}}$ remains of a similar quality. This can be inferred from the fact that, for increasing $N_{\text{max}}$ parameter, the results become gradually independent of the HO energy quanta.

{In order to extract the extrapolated ground-state energy $E_\infty$, {we adopt the simple} ``Extrapolation B'' of Ref. }\cite{Maris09}{, based on the phenomenological relation
\begin{equation}
E(N_{\text{max}},\hbar\Omega) = A(\hbar\Omega)\exp[-c(\hbar\Omega)N_{\text{max}}]+E_\infty(\hbar\Omega)\,. \label{exponential}
\end{equation}
{Given} the results for $N_{\text{max}}=\lbrace N_{\text{max}}^*-2,\,N_{\text{max}}^*,\,N_{\text{max}}^*+2\rbrace$, one can easily solve
\begin{equation}
E_\infty(\hbar\Omega) \!=\! \frac{E^2(N_{\text{max}}^*,\hbar\Omega)\!-\!E(N_{\text{max}}^*\!-\!2,\hbar\Omega)\,E(N_{\text{max}}^*\!+\!2,\hbar\Omega)}{2E(N_{\text{max}}^*,\hbar\Omega)\!-\!E(N_{\text{max}}^*\!-\!2,\hbar\Omega)\!-\!E(N_{\text{max}}^*\!+\!2,\hbar\Omega)}. \label{Einf}
\end{equation}
In our extrapolations, we have taken $N_{\text{max}}^*=14$ (for $^3$H and $^4$He) and $N_{\text{max}}^*=12$ (for $^6$He). We {have observed} that the extrapolated {ground-state} energies display a reasonable $\hbar\Omega$-independence in the intermediate region $\hbar\Omega=(30\sim50)\,\text{MeV}$. 
To be specific, we choose $E_\infty(\hbar\Omega)$ such that the difference $E(N_{\text{max}}^*+2,\hbar\Omega)-E_\infty(\hbar\Omega)$ is minimized \citep{Maris09}. 

The uncertainties in $E_\infty(\hbar\Omega)$, depicted as {horizontal} bands in Fig. \ref{fig:1}, have been determined through applying Eq. \eqref{Einf} with  {$N_{\text{max}}^*=12$} (for $^3$H and $^4$He) and {$N_{\text{max}}^*=10$} (for $^6$He). Still, there is the caveat that the upper uncertainty of an extrapolated result should not extend higher than the result for the largest $N_{\text{max}}$ in consideration; note that the variational principle guarantees that, for any finite truncation of the model space, each eigenvalue provides an upper bound for the ground-state energy in the full model space. {We apply such principle in its strong (or global) version, meaning that our extrapolated energy can never lie above the minimum of the largest-$N_{\text{max}}$ curve. This said, one also needs to be careful} to not simply choose the error associated to the optimal $\hbar\Omega$ from which the corresponding central value was extracted. This is due to the fact that the extrapolated results from two consecutive sets of $N_{\text{max}}$ tend to cross in the close neighborhood of such optimal $\hbar\Omega$, thus leading to underestimated uncertainties when a small $\hbar\Omega$ step is used. Instead, we examined the uncertainties around the optimal $\hbar\Omega$ value (mostly above it) {in order to obtain} such uncertainties. 

We summarize our results in Table \ref{tab1}, showing that the overbinding is {reduced by} about $70\%$ for the three nuclei. However, note that for both potentials studied here, {we find} that $^6$He is \textit{above} the $^4$He threshold. This is an inconvenience that appears with other interactions though, see \textit{e.g.} Ref. \citep{Pudliner:1997ck} and the beyond-LO results of Ref. \citep{LENPIC18}. Yet, the Daejeon16 and JISP16 $NN$ interactions do succeed in amending such an issue, see \textit{e.g.} Ref. \citep{Shirokov16} and references therein.

\begin{figure*}[t!]
\begin{center}
\includegraphics[scale=.31]{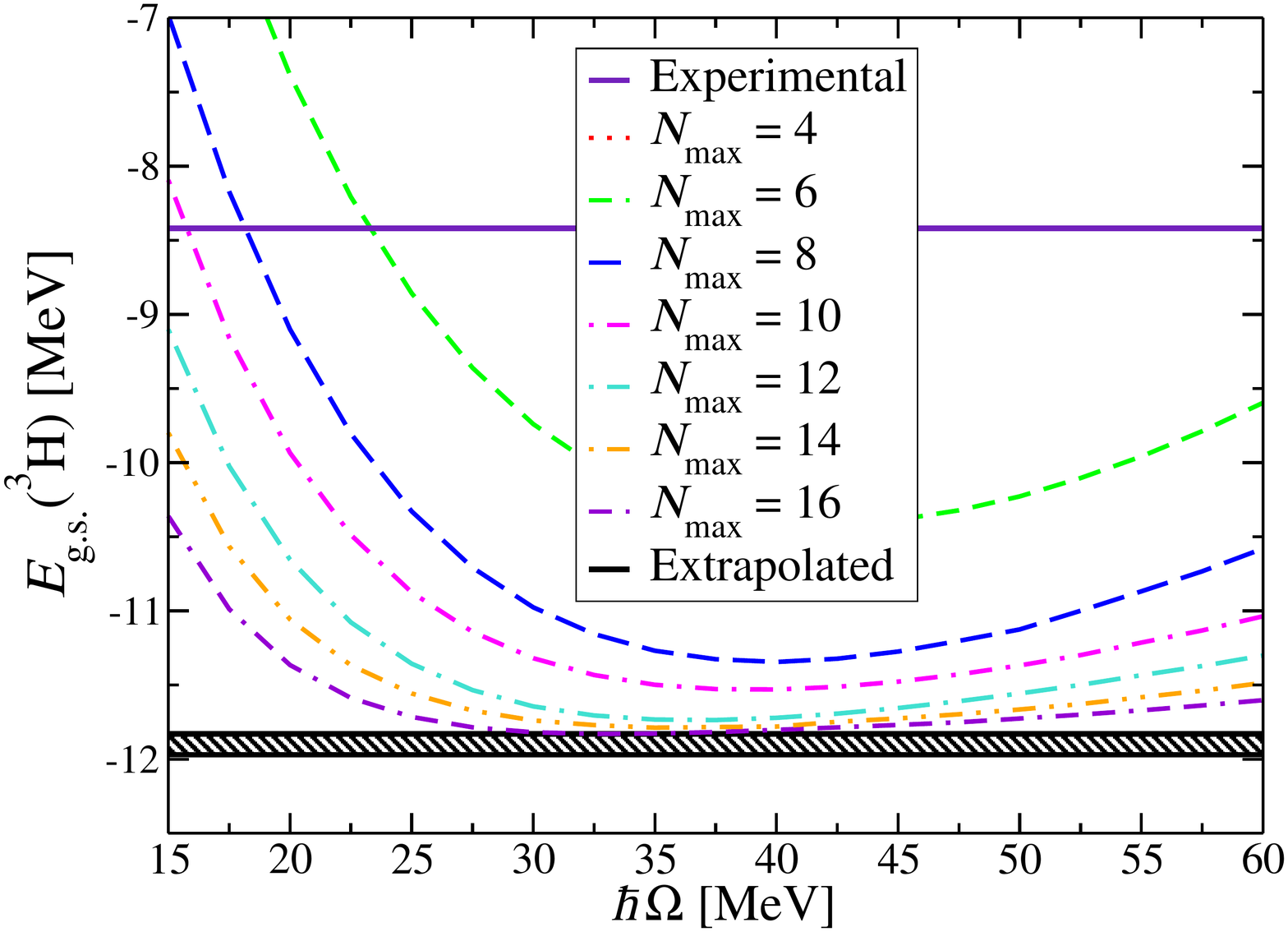}\includegraphics[scale=.31]{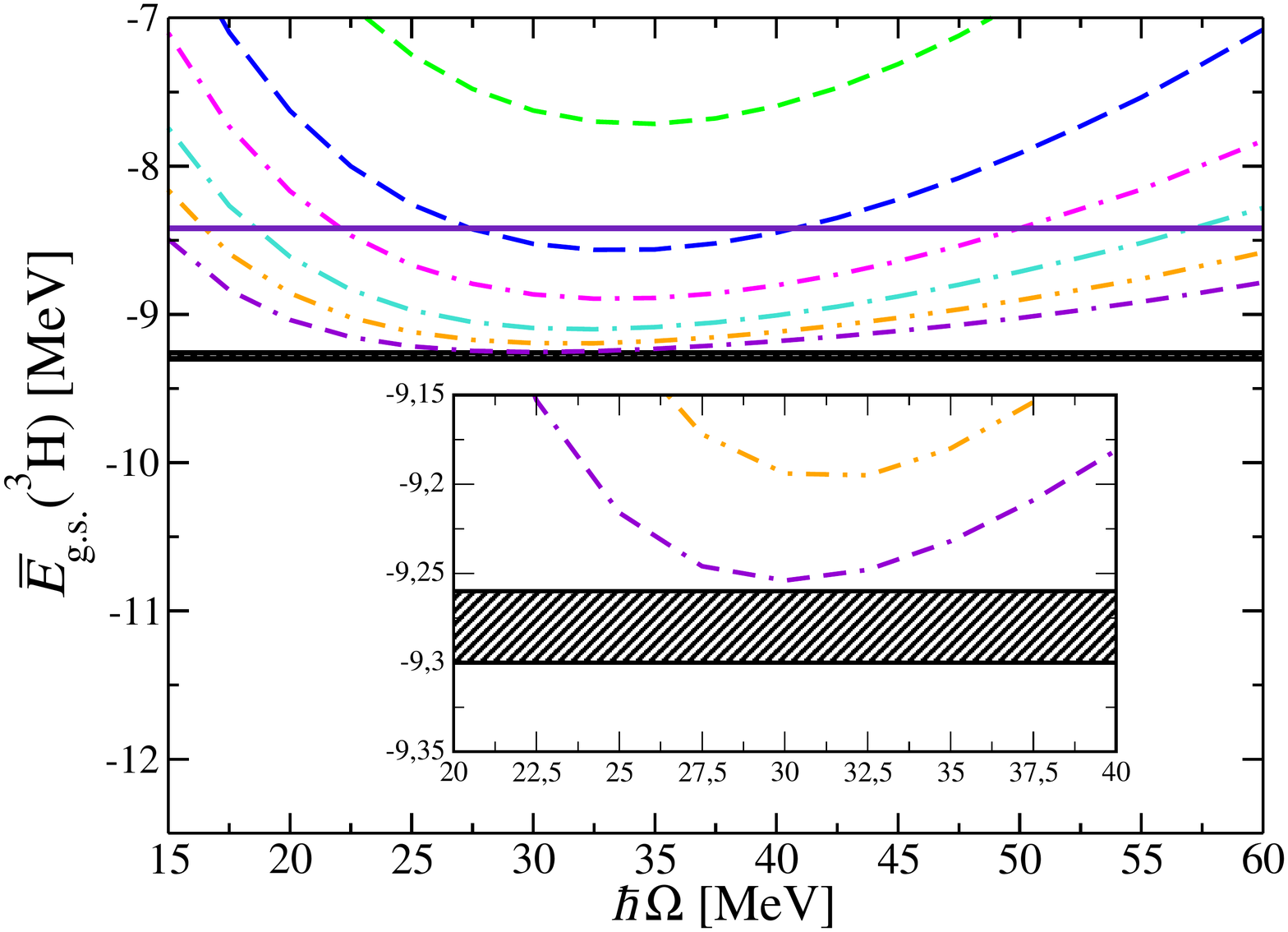}

\bigskip
\includegraphics[scale=.31]{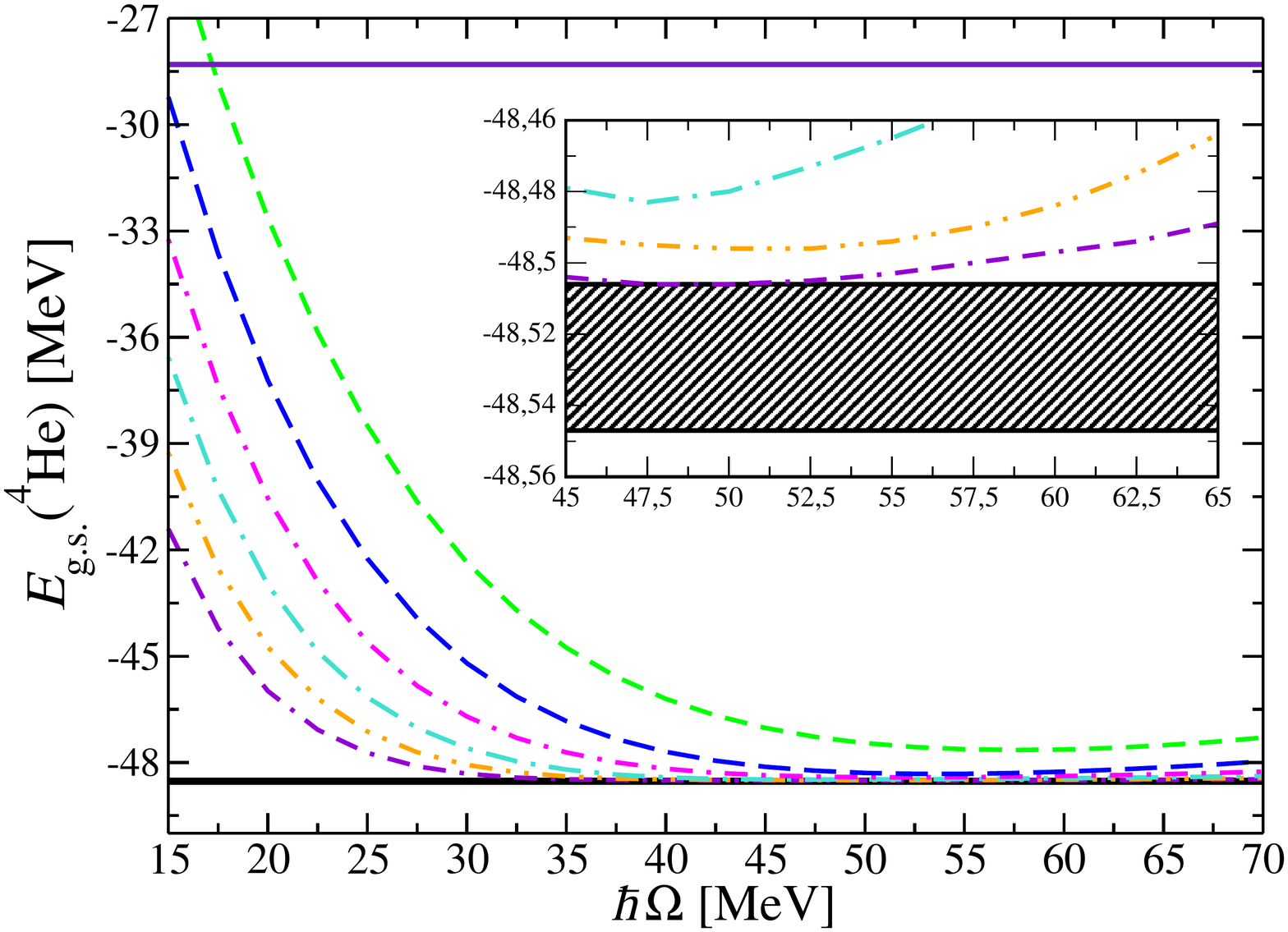}\includegraphics[scale=.31]{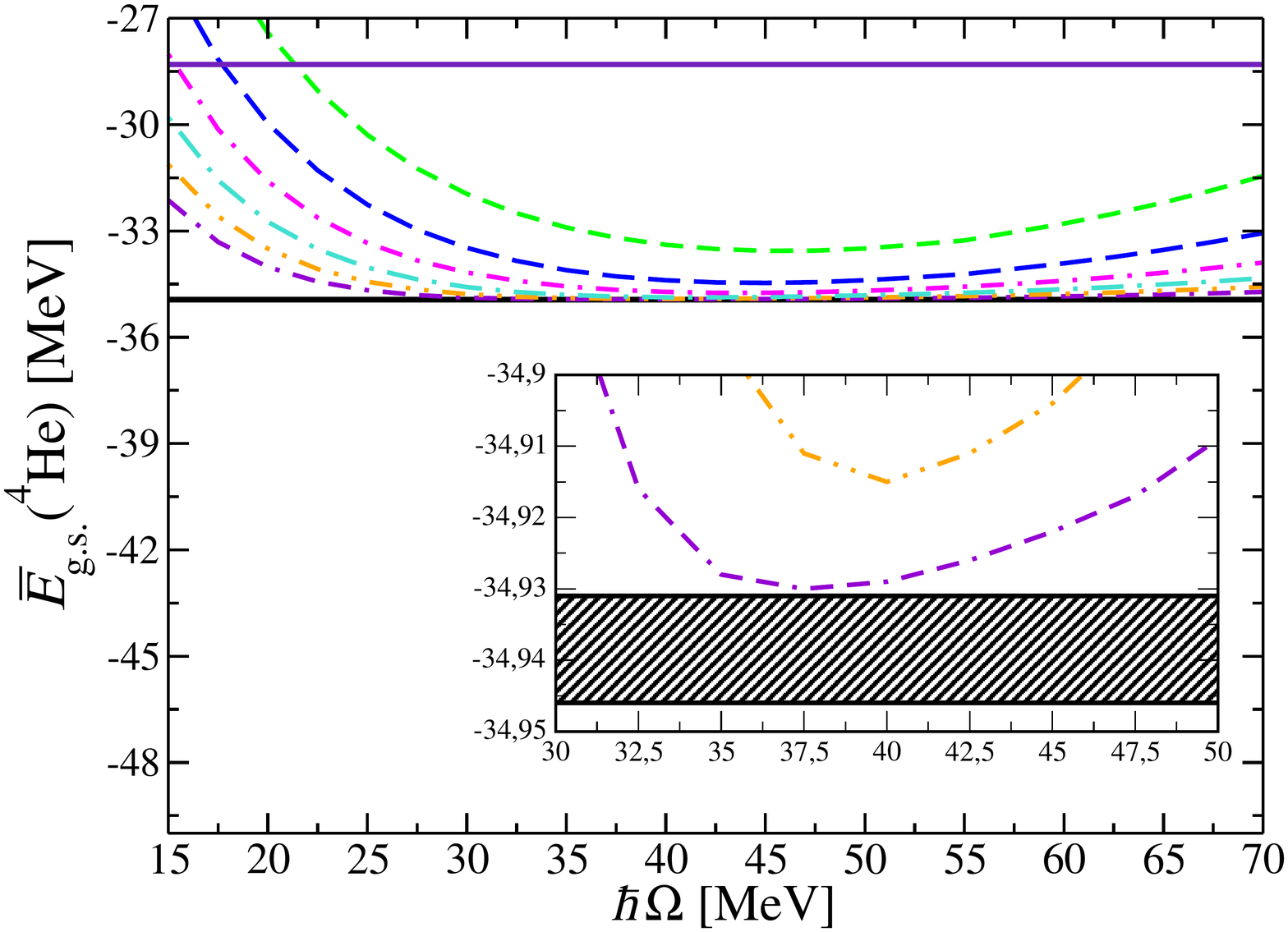}

\bigskip
\includegraphics[scale=.31]{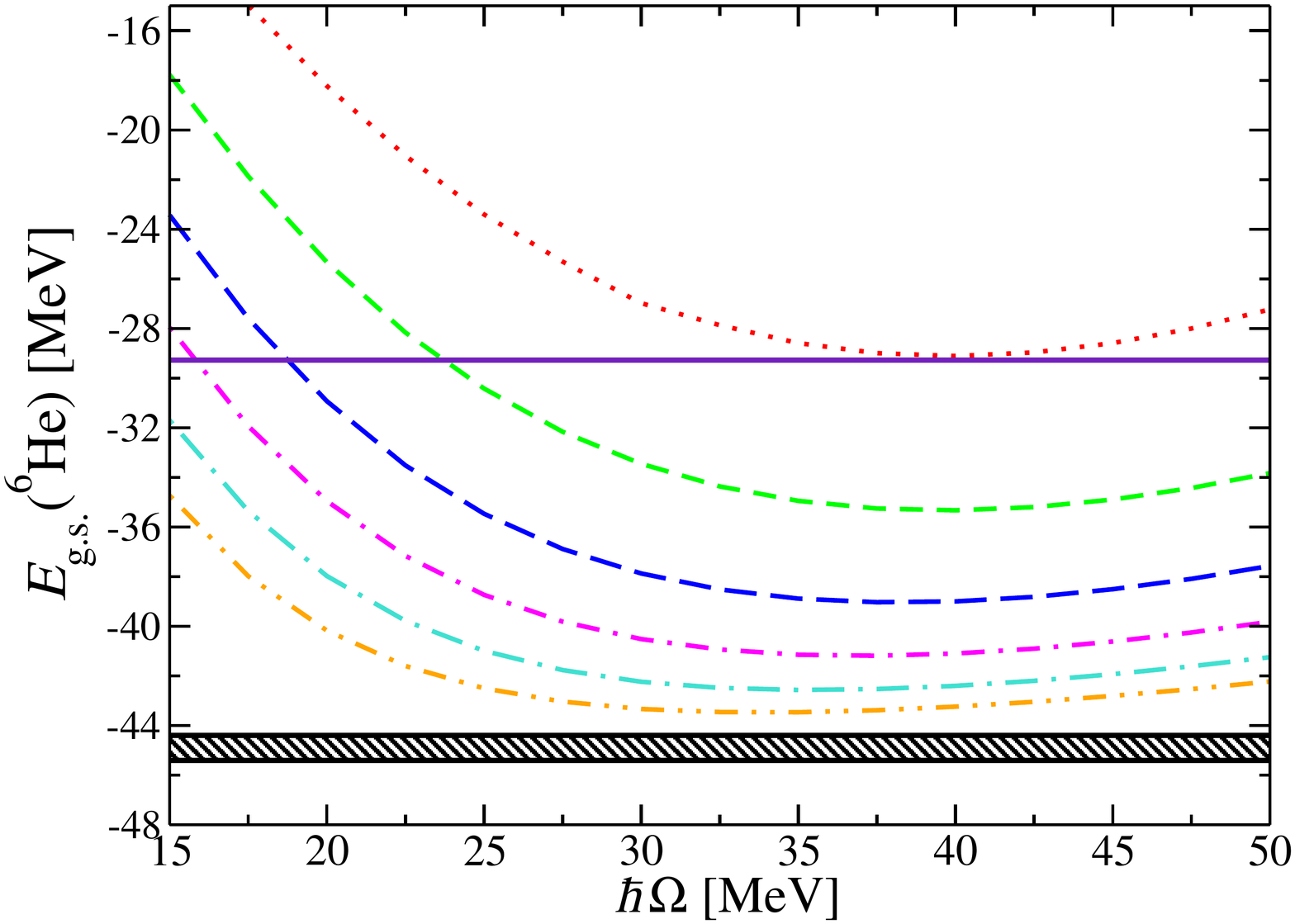}\includegraphics[scale=.31]{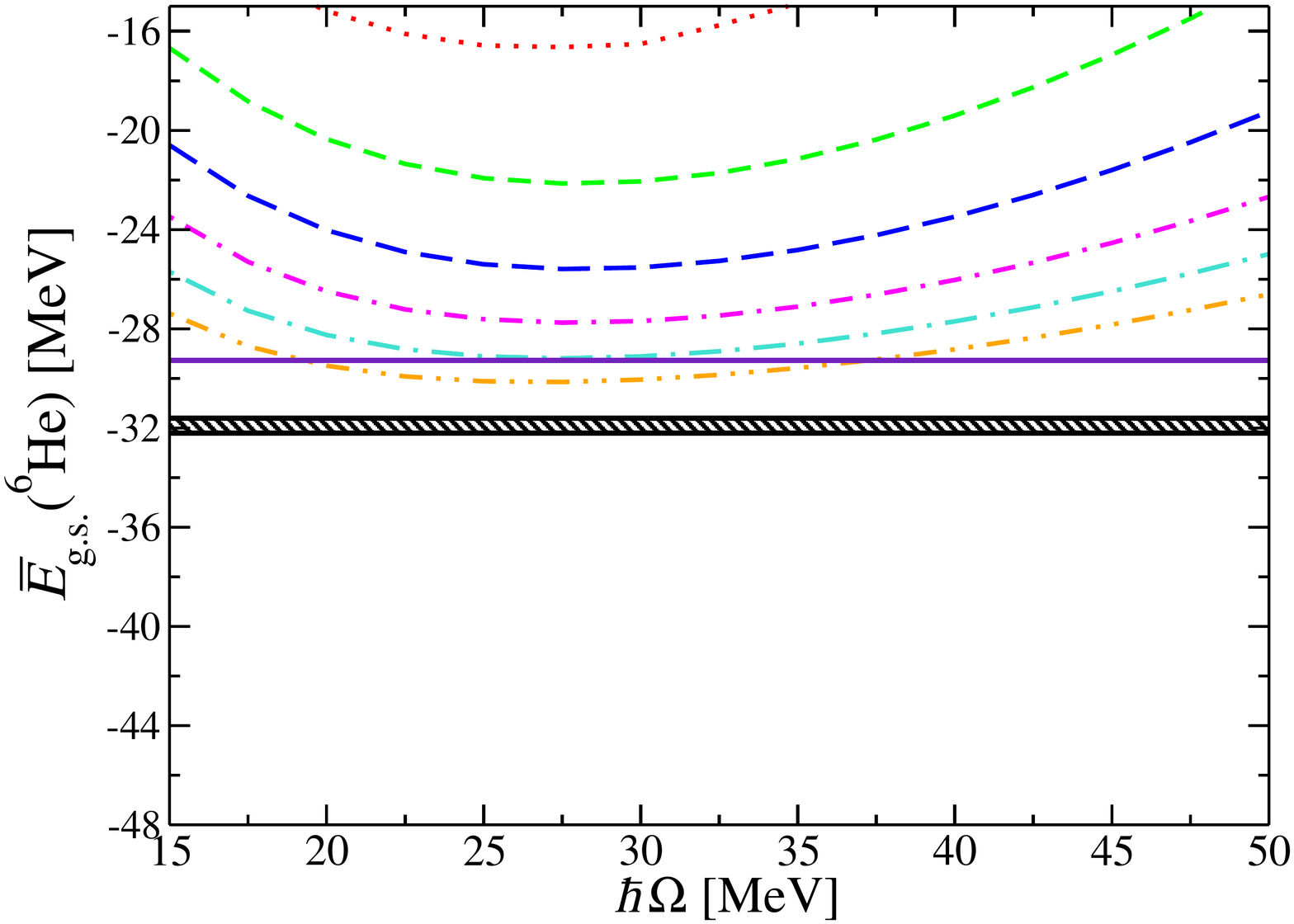}
\end{center}

\caption{Ground-state energies {$E_{\text{g.s.}}$ and $\overline{E}_{\text{g.s.}}$} of $^3$H, $^4$He and $^6$He with \textit{LENPIC$^{[0]}$} \eqref{VchiW} \citep{Maris:2016wrd,LENPIC18,Epelbaum:2018ogq,Semilocal} and \textit{modified LENPIC$^{[0]}$} \eqref{VchiPhi} interactions {respectively}, with the regularization prescription of Eqs. \eqref{short} and \eqref{long} and a coordinate cutoff $R=0.9\,\text{fm}$, together with the corresponding experimental values.  
For both potentials, the infinite-$N_{\text{max}}$ extrapolated results and the associated uncertainties shown by black bands have been obtained through Eq. \eqref{Einf} (see the main text for further explanations).}
\label{fig:1}
\end{figure*}

\begin{table}[ht]
\caption{\label{tab1}
Extrapolated ground-state energies $E_{\text{g.s.}}$ and $\overline{E}_{\text{g.s.}}$ of $^3$H, $^4$He, $^6$He with \textit{LENPIC$^{[0]}$} \citep{Maris:2016wrd,LENPIC18,Epelbaum:2018ogq,Semilocal} and \textit{modified LENPIC$^{[0]}$} interactions respectively. {Note that the asymmetric character of some intervals of confidence is due to the suppression of the positive error by the variational principle. Our results for $E_{\text{g.s.}}(^3\text{H})$ and $E_{\text{g.s.}}(^4\text{He})$ are to be compared with the ones reported in Ref. \citep{LENPIC18} ($-11.747$ and $-48.39$ MeV, respectively), where 
the charge dependence of the $NN$ interaction is explicitly taken into account, allowing us to conclude that its effect is small for these nuclei.}
}
\begin{center}
\begin{tabular}{| c | c | c | c |}
      \hline\hline      \multicolumn{4}{|c|}{\textsc{Ground-state energies}} \\
 \hline            \,\,\,\,\, Nucleus \,\,\,\,\,& \,\,\,\,\,$E_{\text{g.s}}\,[\text{MeV}]$\,\,\,\,\, & \,\,\,\,\,$\overline{E}_{\text{g.s}}\,[\text{MeV}]$\,\,\,\,\, & \,\,\,\,\,$E_{\text{g.s}}^{(\text{exp})}\,[\text{MeV}]$\,\,\,\,\, \\ 
      \hline &&&\\
      $^3$H & {$-11.87^{\,+0.04}_{\,-0.10}$} & {$-9.28 \pm 0.02$}   & $-8.42$\\ &&&\\
      $^4$He & {$-48.507^{\,+0.001}_{\,-0.040}$} & {$-34.936^{\,+0.005}_{\,-0.010}$}  & $-28.30$\\ &&&\\
      $^6$He & {$-44.9 \pm 0.5$} & {$-31.9 \pm 0.3$} & $-29.27$\\ &&&\\
      \hline\hline
   \end{tabular}
\end{center} 
\end{table}

\subsection{Point-proton RMS radii}

Our results for the point-proton radii of $^3$H, $^4$He and $^6$He are plotted in Fig. \ref{fig:2}. Unlike the energy, which is sensitive to the intermediate and short-range correlations, the RMS-radius operator is a long-range operator. In general, long-range operators display a poorer convergence with $N_{\text{max}}$. This is due to the fact that the HO eigenfunctions fall asymptotically as $e^{-r^2/(2b^2)}$ \eqref{eigenfunctions}, while bound-state functions actually fall as $e^{-\kappa r}${. Hence,} increasing the size of the model space increases the radial {extent} of the NCSM ground-state wavefunction, but it does not circumvent its unphysical damping with respect to the true wavefunction. Extracting a robust extrapolation of the point-proton RMS radii would thus {require} new developments and/or larger basis spaces. We note that in the literature there are phenomenological prescriptions accounting for the dependence of the radii on the size of the model space analogous to Eq. \eqref{Einf} for the energies, see \textit{e.g.} Ref. \citep{Shin17}. However, in this work we will adopt the prescription of taking as our guess for the estimated radius the crossing point of the $\hbar\Omega$-dependence of the radii obtained with different $N_{\text{max}}$ \citep{Caprio14}; see Table \ref{tab2}. For the three nuclei, we note that the extra repulsion induced by the modified LENPIC$^{[0]}$ interaction {results in} the crossing point {being} shifted to the left (\textit{i.e.} it appears for smaller values of $\hbar\Omega$) with respect to the original LENPIC$^{[0]}$ interaction. {The overall effect of the modified interaction is to produce a point-proton RMS radius that is larger than the one produced by the original interaction, in closer agreement with experiment.}

\begin{figure*}[t!]
\begin{center}
\includegraphics[scale=.31]{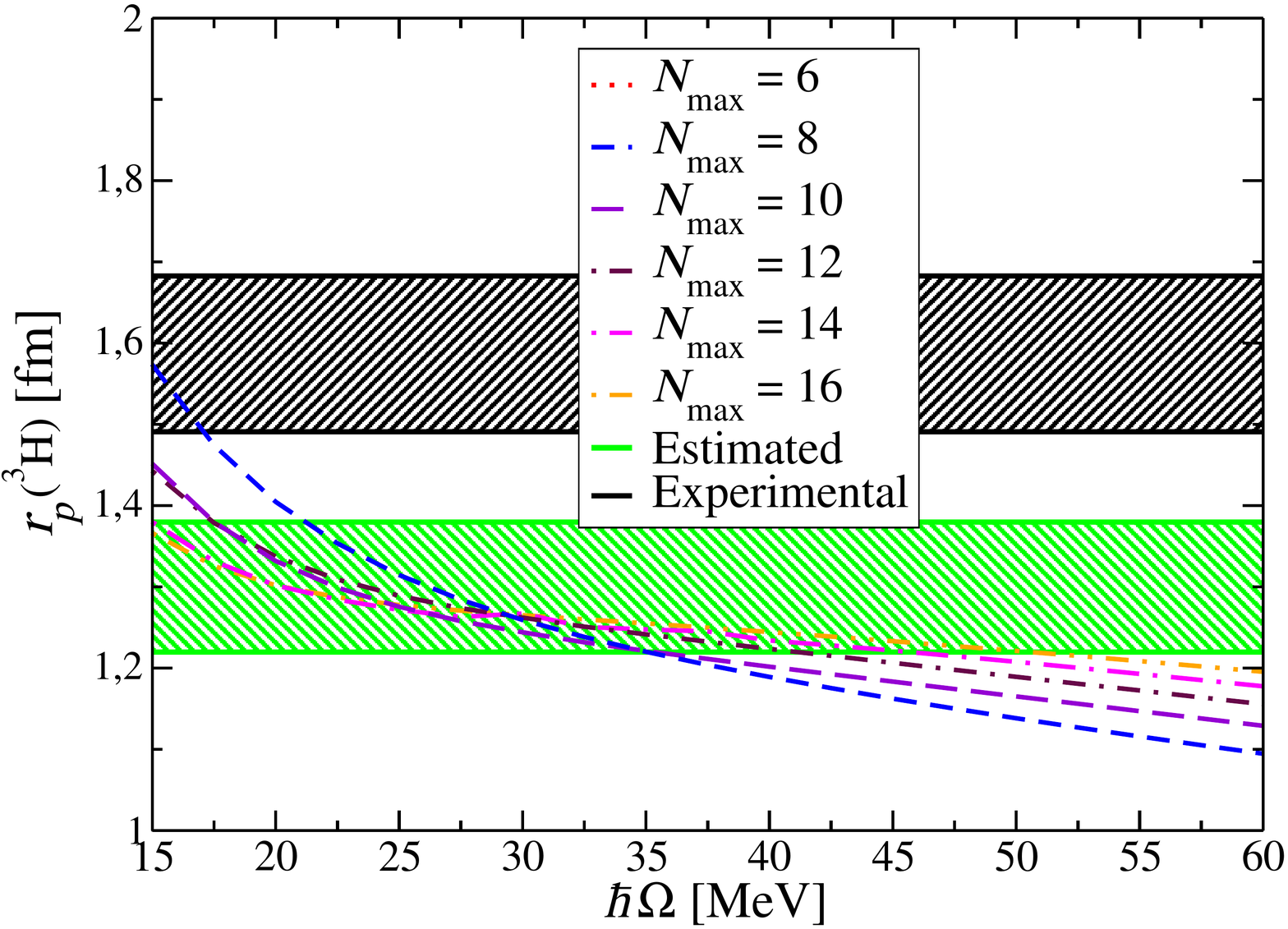}\includegraphics[scale=.31]{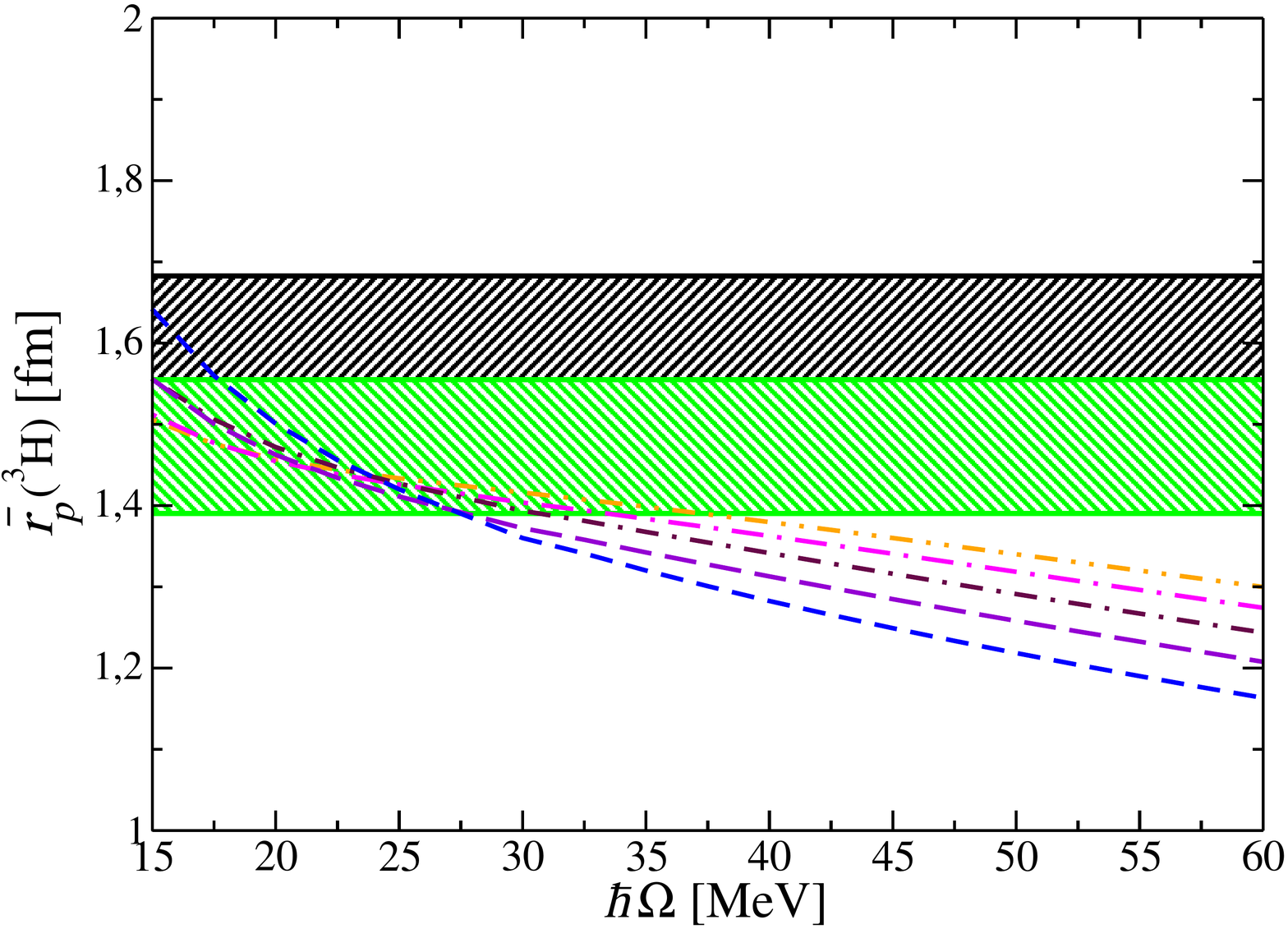}

\bigskip
\includegraphics[scale=.31]{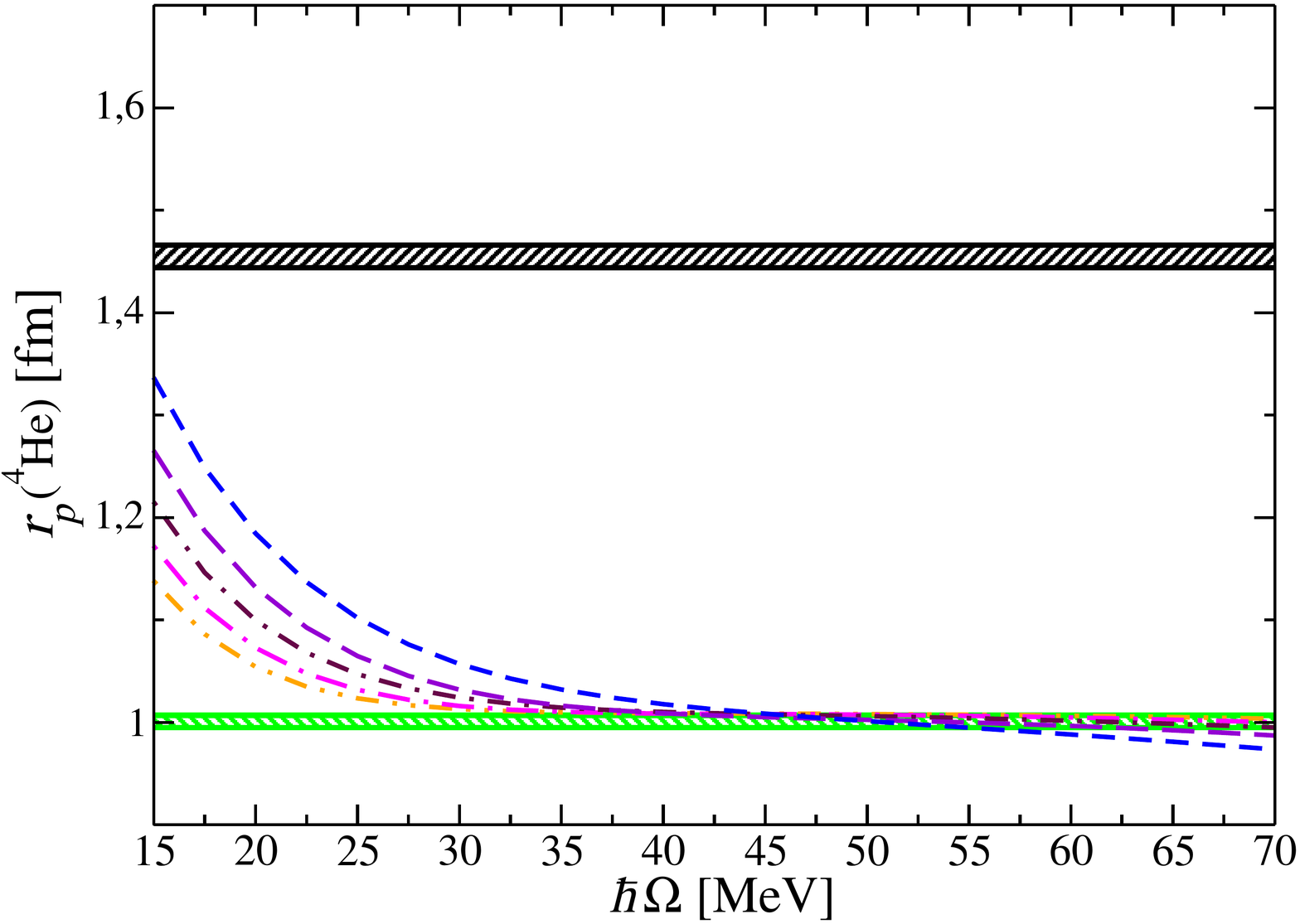}\includegraphics[scale=.31]{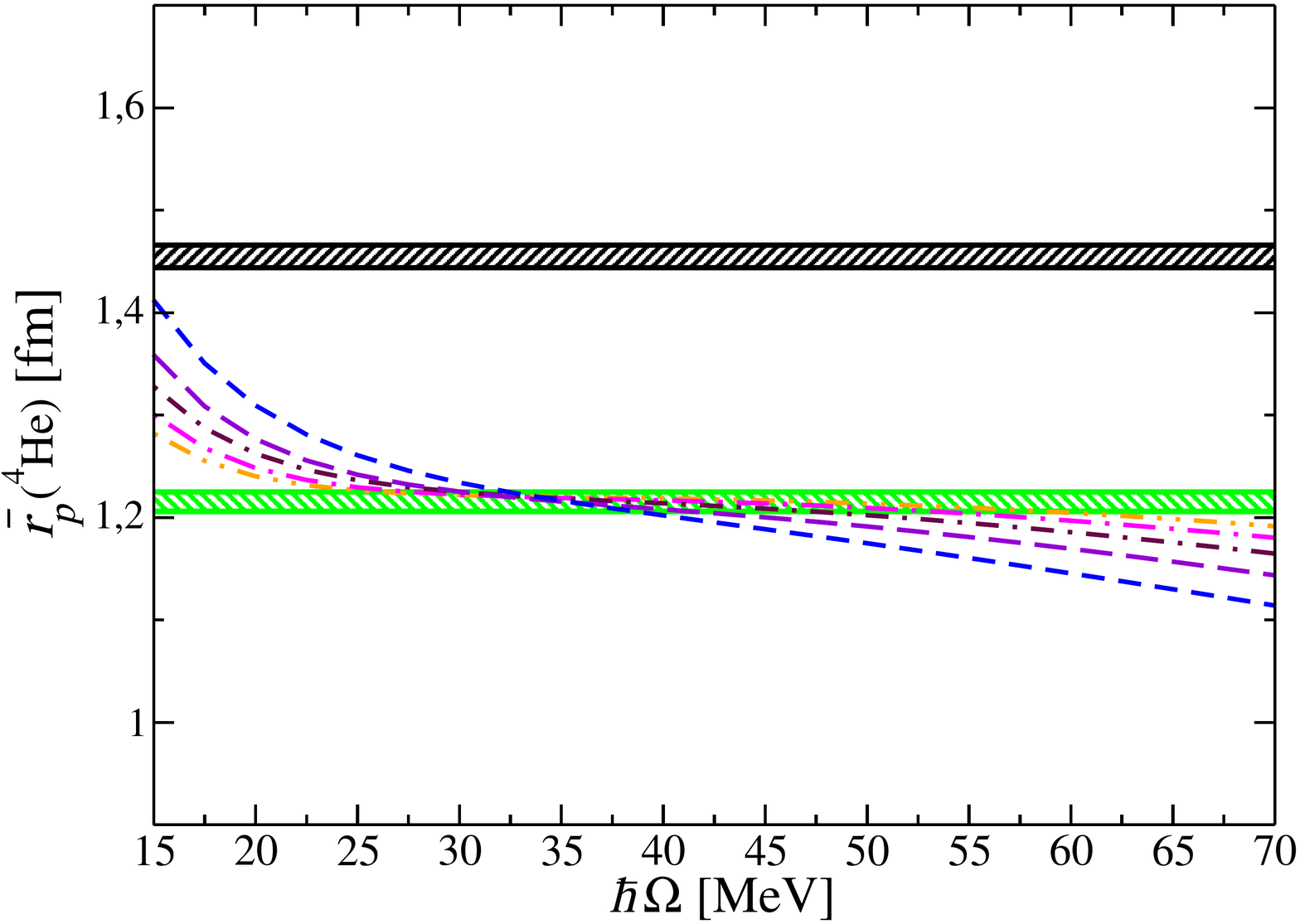}

\bigskip
\includegraphics[scale=.31]{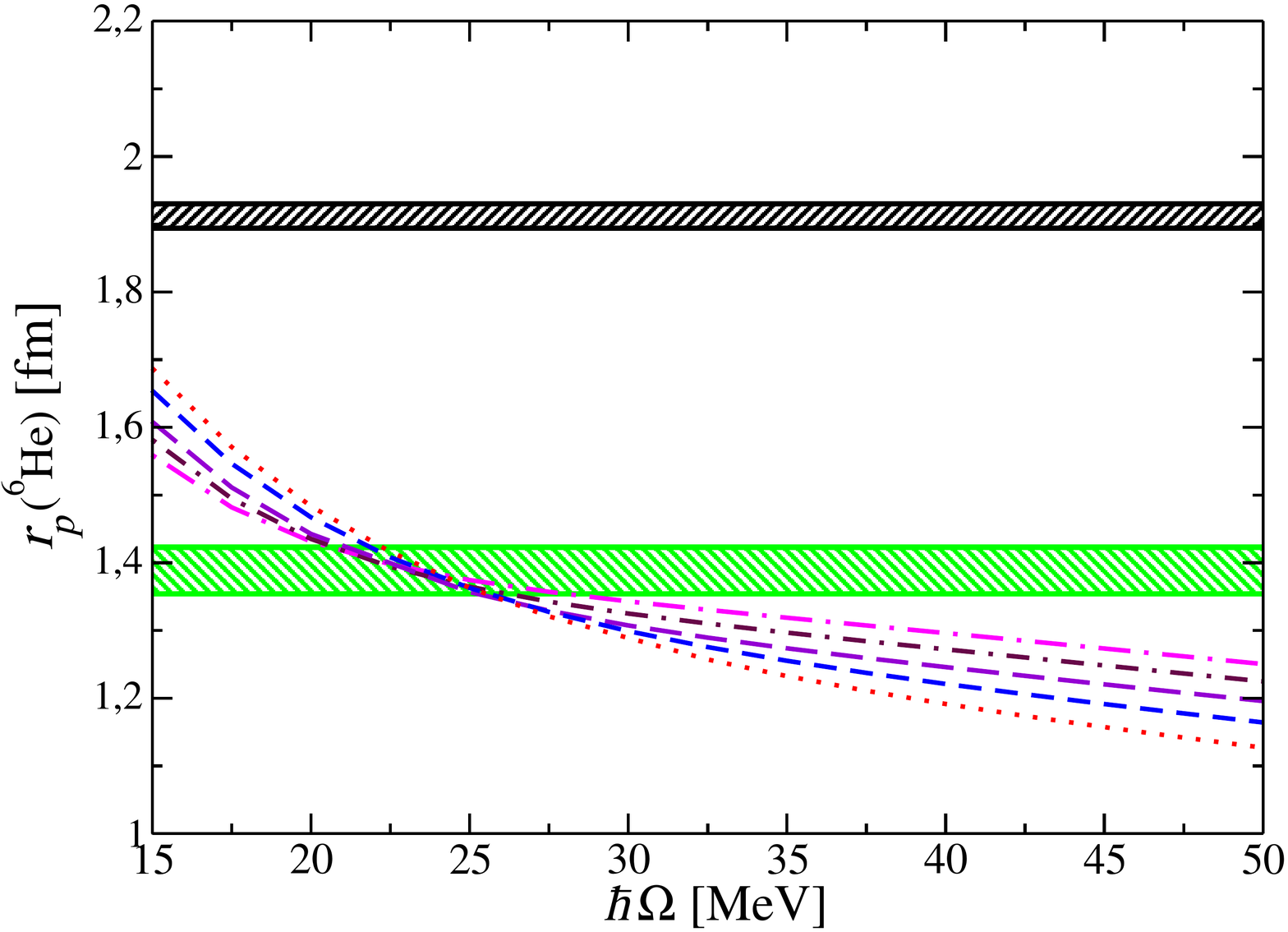}\includegraphics[scale=.31]{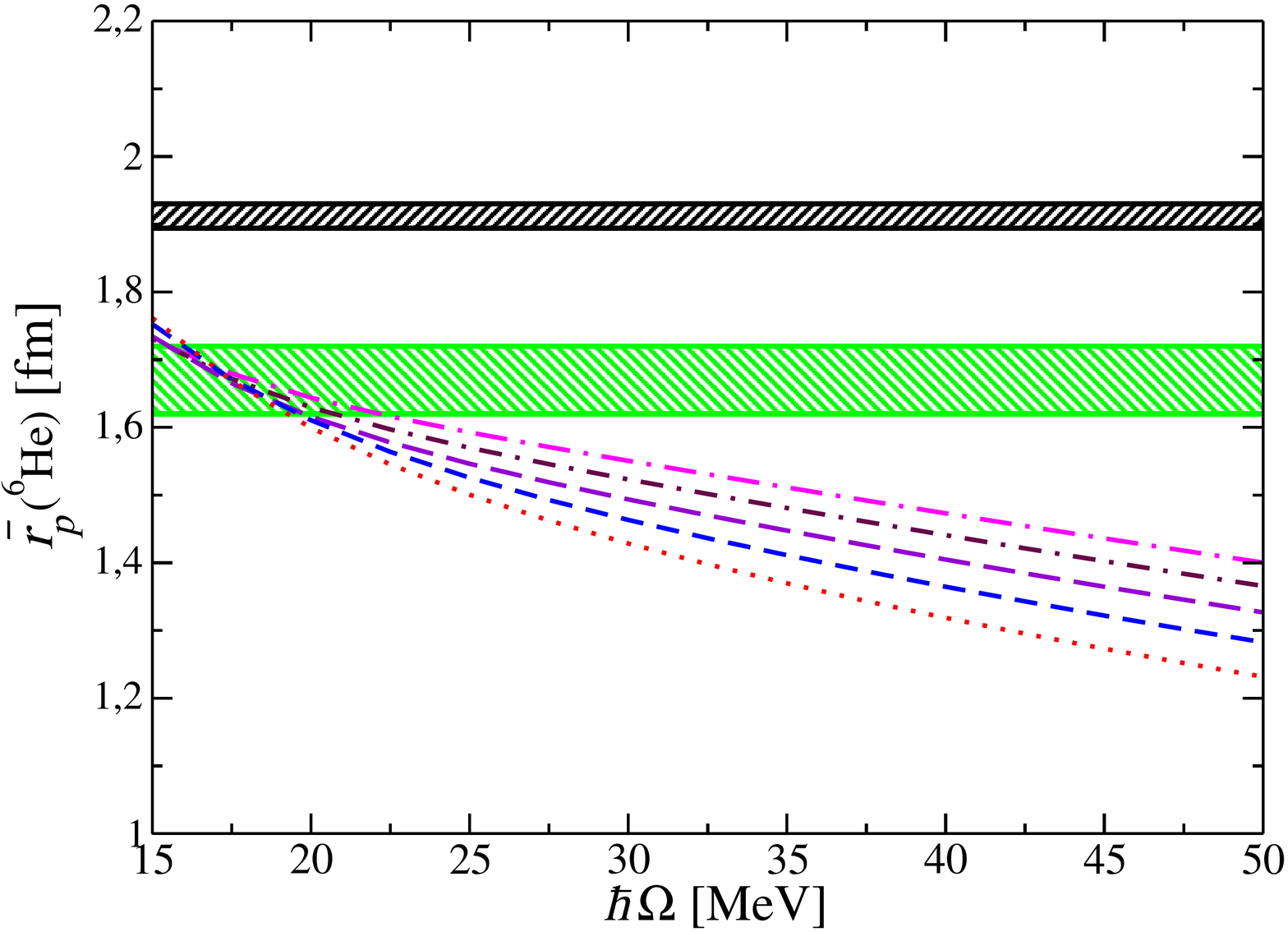}
\end{center}

\caption{Point-proton RMS radii {$r_{\text{p}}$ and $\overline{r}_p$} of $^3$H, $^4$He and $^6$He with \textit{LENPIC$^{[0]}$} \eqref{VchiW} \citep{Maris:2016wrd,LENPIC18,Epelbaum:2018ogq} and \textit{modified LENPIC$^{[0]}$} \eqref{VchiPhi} interactions {respectively}, with the regularization prescription of Eqs. \eqref{short} and \eqref{long} and a coordinate cutoff $R=0.9\,\text{fm}$. 
The dashed black (green) band represents the experimental error (the uncertainty of our estimated result). {Note that the upper (lower) limit of the latter was obtained from the highest (lowest) point where two curves with different $N_{\text{max}}$ cross each other.}}
\label{fig:2}
\end{figure*}

\begin{table}[b!]
\caption{\label{tab2}
Point-proton RMS radii $r_{p}$ and $\overline{r}_{p}$ of $^3$H, $^4$He, $^6$He with \textit{LENPIC$^{[0]}$} \citep{Maris:2016wrd,LENPIC18,Epelbaum:2018ogq,Semilocal} and \textit{modified LENPIC$^{[0]}$} interactions respectively. {Note that the ``experimental'' point-proton radii are extracted from the experimental charge radii given in Refs. }\citep{Amroun94,Wang04}.}
\begin{center}
\begin{tabular}{| c | c | c | c | }
      \hline\hline      \multicolumn{4}{|c|}{\textsc{Point-proton radii}} \\
 \hline    \,\,\,\,\, Nucleus \,\,\,\,\, & \,\,\,\,\, $r_p\,[\text{fm}]$ \,\,\,\,\, & \,\,\,\,\, $\overline{r}_{p}\,[\text{fm}]$ \,\,\,\,\, & \,\,\,\,\, $r_{p}^{(\text{exp})}\,[\text{fm}]$ \,\,\,\,\, \\ 
      \hline &&&\\
       $^3$H &$1.3\pm0.1$ &$1.5\pm0.1$ & $1.587\pm0.096$ \\ & & & \\
      $^4$He &$0.99\pm0.01$ &$1.22\pm0.01$ & $1.455\pm0.011$ \\ & & &\\
      $^6$He &$1.39\pm0.05$ &$1.67\pm0.05$ & $1.912 \pm 0.018$  \\ & & &\\
      \hline\hline
   \end{tabular}
\end{center}
\end{table}

{We see that the modification of the LO chiral potential described in Sec. \ref{SecI} allowed us to produce new LO results for two nuclear magnitudes, namely the ground-state energy and the point-proton RMS radius, of three positive-parity light nuclei. These predictions are closer to experiment than the LO results of Ref. \citep{LENPIC18} obtained under the assumption of NDA. In particular, we notice that the excess of $NN$ attraction anticipated by Weinberg power counting at LO generates, for these three nuclei, an overestimation of the binding energies and an underestimation of the radii that can be both easily and significantly improved with our proposal.}

\section{Conclusions and perspectives \label{SecIII}}

In this work we have explored the consequences in the description of light nuclei {produced by a} rearrangement of the short-range part of {a} two-nucleon chiral potential that is most commonly employed in current \textit{ab initio} calculations of light nuclei, \textit{i.e.} the one grounded on the Weinberg power counting. {We} have followed Ref. \citep{zero} in promoting subleading (repulsive) interaction terms that capture, already at leading order in the effective expansion, the zero of the $^1S_0$ partial-wave amplitude {--- a zero that appears at a relatively soft scattering momentum according to experiment}. We remark that the proposal here relies merely on the treatment of such momentum as a low-energy scale{. Furthermore,} the $^1S_0$ channel amplitude is unique in displaying such a low-lying zero {along with} a very shallow pole{. We distinguish the situation in the $^1S_0$ channel} from the $^3S_1$ channel (where the amplitude zero lies beyond the pion-production threshold) and from the $^3P_0$ amplitude ({which} turns around at a lower energy but contains no low-energy pole){. The scattering in these latter two channels} is qualitatively well captured{ }at leading order --- either by the Weinberg scheme for cutoffs below the breakdown scale of the theory \citep{Entem03,Epelbaum05} or for a wider cutoff range through certain renormalization-consistent modifications of the Weinberg prescription \cite{NTvK,Valderrama:2009ei,Valderrama:2011mv,Long:2011qx,Long:2011xw}.

We have shown that the minimal change in the power counting invoked in this work significantly helps to improve the leading-order results, at the level of the ground-state{ }energies and point-proton radii, of the $^3$H, $^4$He and $^6$He nuclei. In particular, the excess of attraction of the $^1S_0$ two-nucleon interaction brought by the Weinberg scheme at leading order yields a pattern of overbinding in the {ground-state} energies and underprediction of the radii for those nuclei \citep{LENPIC18}. {We show that this deficiency} can be addressed by the inclusion of repulsive terms as leading-order effects. This may {also} help to alleviate the pressure on higher orders of the effective expansion, thus opening a new avenue {for potentially improved convergence with respect to increasing chiral order}.

Of course, before a claim of convergence of this rearrangement can be made, one needs to produce results beyond leading order in the expansion{ }for observables{. We} intend to pursue this task in future work. {{We also need to study} how charge-dependent and charge-asymmetric terms should be encoded in our proposed potential expansion, in the spirit of what was done in Ref. }\citep{Bira95}{ for Weinberg's power counting.}

{The path forward has significant uncertainties. Note that} {--- in spite of some exceptions, such as the recent work of Ref. \citep{Yang19} ---} the general way to proceed in current \textit{ab initio} calculations is, following Weinberg's original idea \citep{Weinberg90,Weinberg91}, to treat subleading terms of the potential on the same footing as its leading part (\textit{i.e.} non-perturbatively){. However,} some authors \citep{NTvK,Birse:2005um,Valderrama:2009ei,Valderrama:2011mv,Long:2011qx,Long:2011xw,Long:2012ve,Song:2016} argue that, in order not to undermine cutoff independence of observables, such subleading contributions need to be added as perturbations on top of the infinitely iterated leading-order potential. {Still, other} authors \citep{Epelbaum09,Epelbaum13,Epelbaum18,Epelbaum20}{ }disagree with this conception, as they claim that cutoff dependence of observables is guaranteed to be reasonably mild provided that one sticks to cutoff values that are softer than the breakdown scale of the EFT --- typically, below (500 $\sim$ 600) MeV in \textit{ab initio} calculations.

{It is worth recalling that} cutoff-convergence issues of the Weinberg counting already emerge in the two-nucleon sector at leading order itself, as first noticed in Ref. \citep{NTvK}. {Such issues arise} in those channels where one-pion{ }exchange{ }--- which is prescribed to be leading order under the assumption of naive dimensional analysis --- is both singular and attractive{. That is, in light of the $1/r^3$ divergence of the potential at small $r$,} the leading-order amplitude does not converge for large enough cutoffs if no repulsive contact term is employed. However, in Weinberg power counting a contact term affecting a partial wave with orbital angular momentum $\ell$ is prescribed to contribute no less than $2\ell$ orders down with respect to leading order. Yet, one can easily check that the $^3P_0$ amplitude becomes ill-defined when cutoffs in the range $500\,\text{MeV} \sim 1\,\text{GeV}$ are used unless an unexpected contact term with two derivatives is promoted from next-to-next-to-leading order to leading order. But, since there is an infinite number of partial waves where one-pion exchange is both singular and attractive, Refs. \cite{NTvK,Valderrama:2009ei,Valderrama:2011mv,Long:2011qx,Long:2011xw} advocate to treat one-pion exchange as a subleading (perturbative) correction in those channels where the centrifugal barrier becomes effective. In particular, the $^3P_0$ channel might be the only partial wave with $\ell\geqslant1$ where one-pion exchange needs to be retained at leading order, as first pointed out in Ref. \citep{Birse07}; such a hypothesis is backed by the more recent work of Refs. \citep{Wu19,Kaplan19}. Hence, in the future we plan to promote the $^3P_0$ contact term to leading order and see how this may improve the description of heavier nuclei. We remark, however, that such a modification in the Weinberg counting is mainly motivated by renormalization requirements, unlike the one proposed here, which was aimed at improving the agreement with phenomenological evidence. {Finally, we comment that} cutoff dependence of observables was not studied here, but it will be addressed in future work. 

\section*{Acknowledgments}

M.S.S. would like to thank U. van Kolck, B. Long, {I.J. Shin,} and T. Frederico for useful {discussions} and encouragement, particularly at the early stages of the project. We are grateful for hospitality to the Institute for Basic Science (Daejeon) during the NTSE--2018 conference {(M.S.S., N.A.S., A.M.S., J.P.V.)}, to the University of Nanjing during the workshop ``Effective field theories and \textit{ab initio} calculations of nuclei'' (M.S.S., A.M.S., J.P.V.), and to the Iowa State University (M.S.S.), where parts of this work were carried out. {This work is supported in part by the IN2P3/CNRS (France), by the Russian Foundation for Basic Research under Grant No. 20-02-00357, and by the U.S. Department of Energy under Grants No. DESC00018223 (SciDAC/NUCLEI) and No. DE- FG02-87ER40371.} Computational resources were provided by the National Energy Research Scientific Computing Center (NERSC), which is supported by the Office of Science of the U.S. Department of Energy under Contract No. DE-AC02-05CH11231.

\end{document}